\documentclass{elsart}
\usepackage{amsbsy}
\usepackage{amssymb}
\usepackage{epsf}
\begin{document}
\begin{frontmatter}
\title{
Possibility of time reversal symmetry violation at proton
deuteron forward elastic scattering}
\author{S.L. Cherkas }
\address{
Institute of Nuclear Problems, Bobruiskaya 11, Minsk 220050, Belarus,
cherkas@inp.minsk.by}
\begin{abstract}
We consider T-violating P-conserving
proton-deuteron forward elastic scattering
amplitude in a frame
of the Glauber multiscattering theory,
proceeding  from T-odd, P-even N-N
interaction.
\end{abstract}
\end{frontmatter}


keywwords: time-reversal, symmetry

\everymath {\displaystyle}
\section{Introduction}

Discovery
of CP symmetry breaking
in K-meson decay
stimulated a search of T-
and CP- non-invariant interactions in other systems.
It is important to
distinguish T- P- breaking interaction from T-breaking  P-conserving one.
While T-odd  P-odd interaction naturally arises in the Standard model
through CP-breaking phase of the Kobayashi-Maskawa matrix or in QCD
through $\theta$  term,
introduction of
T-breaking P-conserving interactions has
no natural implementation
on a quark level.
The investigation of P-even T-odd interaction (TRV) of non-weak origin
continue to enjoy popularity in nuclear and nucleon-nucleon
physics.
Note, that the T-breaking, P-conserving interaction arising from
interference of
T- P- breaking interaction with P-odd weak one, should be very small
and can not be considered.

Null test for the T-odd P-even
interaction is finding of the
\begin{equation}
F_T(0)={\mathfrak F}_T
\bigl\{(\boldsymbol \sigma \cdot {\boldsymbol S} \times {\boldsymbol n})
({\boldsymbol S} \cdot {\boldsymbol n})
+({\boldsymbol S} \cdot {\boldsymbol n})
(\boldsymbol \sigma \cdot {\boldsymbol S}
\times {\boldsymbol n})\bigr\}
\end{equation}
term \cite{bar1,bar2,ex} in the forward elastic
scattering amplitude of the particle with the $1/2$ spin by the nucleus with
the spin $S\ge1$. $ {\boldsymbol S} $ is the  nucleus spin operator, $
{\boldsymbol n} $ is the unit vector in the direction of the projectile
particle momentum, $\boldsymbol \sigma$ is the Pauli matrix of the nucleon spin.

The heavy nuclei exhibit the large enhancement of P-odd effects
but for T-odd P-even interactions the enhancement can be
suppressed \cite{Haxt}.
So, the advantage of  heavy nuclei compared to light is questionable.
On the other hand,
the proton-deuteron scattering considered here fully utilise a high
luminosity of proton beams of COSY.
Such an experiment is planed to be performed
by TRV collaboration \cite{TRV}
as an
internal target experiment in the cooler synchrotron COSY.

We consider the T-violating P-conserving
p-d forward scattering amplitude in
the frame of the Glauber multiscattering theory proceeding from
T-odd, P-even terms of N-N interaction. The N-N forward elastic
scattering amplitude does not contain T-odd, P-even terms:
such a term exists in scattering amplitude only at non-zero angles.
The mechanism of  $\mathfrak F_T$-term arising is double
proton scattering by deuteron nucleons. In the first
collision the proton is scattered by TRV interaction at a small angle
$\boldsymbol \theta$ and in the second collision it is scattered
by T-even interaction at the angle $-\boldsymbol\theta$, as a result,
the T-odd, P-even term appears at the forward $p-d$ elastic
scattering amplitude. The problem is discussed
earlier \cite{Bey}, but
we now take into account the
deuteron non sphericity and calculate the contribution
of the T-odd impurity at the deuteron density matrix.

\section{
Eikonal approximation for particles with spin}

The Glauber multiscattering   theory
can be generalized to the case of particles with spins
\cite{glau,tar}. Let us
consider the small-angle elastic scattering
of the particle with spin by the
$N$
scatterers also with spins and fixed at the points
with the radius vectors ${\boldsymbol r}_\alpha$.
The wave
function of the system $\Psi({\boldsymbol r},{\boldsymbol r}_\alpha)$
satisfies to the
"relativizied" Shredinger equation:
\begin{equation}
\label{sr}
(\nabla^2 + k^2)\Psi({\boldsymbol r},
{\boldsymbol r}_\alpha)=2E V({\boldsymbol r},
{\boldsymbol r}_1
 \cdot \cdot
{\boldsymbol r}_N)\Psi({\boldsymbol r},{\boldsymbol r}_\alpha),
\end{equation}
where $k$ and $E$ are the wave number
and the particle energy accordingly.
The particle interaction with the scatterers
$V({\boldsymbol r},{\boldsymbol r}_1 \cdot \cdot
{\boldsymbol r}_N)$ is an operator
at the particle spin space and
at the every scatterer spin space.
The solution of the equation (\ref{sr}) can be found in the form
$
\Psi({\boldsymbol r})=e^{ikz}\Phi({\boldsymbol r})$.
Substituting this value to the
(\ref{sr}) and neglecting the terms with
second derivative of
$\Phi({\boldsymbol r})$ we come to
\begin{equation}  \label{vz}
\frac{ik}{E}\frac{\partial \Phi({\boldsymbol r})}
{\partial z}= V({\boldsymbol r})\Phi({\boldsymbol r}).
\end{equation}
Equation  (\ref{vz}) is analogous to
to the interaction representation in the
quantum field theory and
consequently it
has solution
$Zexp$ being analogy to the $Texp$:
\begin{equation}
\Phi({\boldsymbol b},z)=Z^\prime exp
\{-\frac{iE}{k}\int_{-\infty}^z
V({\boldsymbol b},z^\prime)dz^\prime\}
\end{equation}
For the scattering amplitude we have:
\begin{equation}
F({\boldsymbol q})=-\frac{E}{2\pi}\int e^{-i
{\boldsymbol q}
{\boldsymbol b}-ikz} V({\boldsymbol r})
\Psi({\boldsymbol r}) d^3{\boldsymbol r}= \frac{ik}{2\pi}\int
e^{-i{\boldsymbol q} {\boldsymbol b}}(1-\Phi({\boldsymbol
b},+\infty))d^2{\boldsymbol b},
\end{equation}
where $\boldsymbol q$ is the momentum transferred.
If $V({\boldsymbol
b},z)=\sum_{\alpha=1}^{N}V_\alpha ({\boldsymbol b} -{\boldsymbol
b}_\alpha,z-z_\alpha)$ and $V_\alpha({\boldsymbol b}
-{\boldsymbol b}_\alpha,z-z_\alpha)$ is
concentrated  near  the point $z_\alpha$
so that the different $V_\alpha$ do not overlap
it is possible to write:
\[
 Z^\prime exp
\{-\frac{iE}{k}\int_{-\infty}^{+\infty}
V(\boldsymbol b,z^\prime)
dz^\prime\}
=\hat Z_\alpha \prod_\alpha Zexp
\left(-\frac{ iE}{k}
\int_{-\infty}^{+\infty}V_\alpha({\boldsymbol b}-
{\boldsymbol b}_\alpha,z-z_\alpha)dz \right).
\]
The operator
$\hat Z_\alpha$ orders the terms in this product
in the direction of
$z_\alpha$ increasing
from the right side to the left one.
By denoting
\begin{equation}
\Gamma_{\alpha}({\boldsymbol b}-
{\boldsymbol b}_\alpha)=1-Zexp \biggl\{ -\frac{iE}{k}
\int_{-\infty}^{+\infty} V_\alpha({\boldsymbol b}-
{\boldsymbol b}_\alpha,z-z_\alpha)dz\biggr\}
\end{equation}
we find
\begin{equation}
F(q)=\frac{ik}{2\pi}\int e^{-i{\boldsymbol q}
{\boldsymbol b}} \langle \mid \biggr\{1- \hat
Z_\alpha\prod_\alpha\left( 1-\Gamma_\alpha ({\boldsymbol b}-
{\boldsymbol b}_\alpha)\right)\biggr\}\mid
\rangle d^2{\boldsymbol b}.
\label{fei}
\end{equation}
$\langle \mid \quad \mid \rangle$
means the averaging over
the displacements and the spin states of the scatterers in the
target nucleus.
 Remember that the
$\hat Z_\alpha$ orders the
terms in the direction of $z_\alpha$
increasing.

\section{ T-odd P-even N-N elastic scattering amplitude}

\begin{figure}[h]
\hspace{2.0cm}
\vspace{0.0cm}
\epsfxsize =7.0cm
\epsfbox[0 0 300 280]{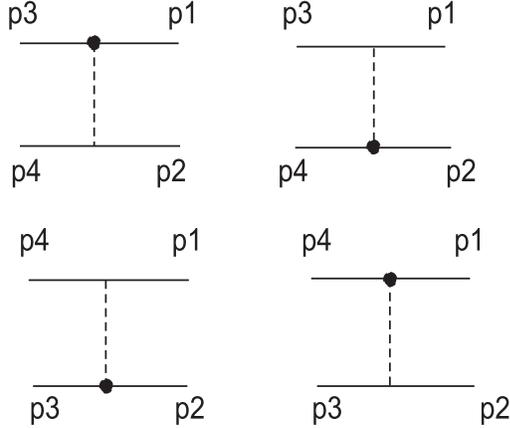}
\vspace{0.0cm}
\caption{One meson exchange diagrams. Black circle
denote T-odd, P-even vertex.}
\label{diag}
\end{figure}

In
view of absence of the fundamental model the T-odd P-even interaction
is usually considered at the nucleon
level in the framework of one meson exchange
mechanism
 which is
responsible for the long range
component of nuclear forces.
$A_1 $ and $ \rho $ are two easiest mesons, exchange of which can
results to the T-odd P-even N-N scattering amplitude \cite{sim}.
The Lagrangian of the
interaction  of  $A_1 $ and $
\rho $ fields with nucleon field can be written down
as \cite{sim}
\begin {eqnarray} L (x) = \bar\psi (x) \{g _ {\rho} \gamma _ {\mu}
{\boldsymbol \tau} \cdot { \boldsymbol \rho} ^ \mu (x) - \bar g_\rho
\frac {g_\rho \kappa} {2m} \sigma _ {\mu\nu} \lbrack {\boldsymbol
\tau} \times \partial ^\nu {\boldsymbol \rho} ^ \mu (x) \rbrack _3
\nonumber \\ +g_A\gamma_\mu\gamma_5\boldsymbol \tau\cdot \boldsymbol
A ^\mu (x) - \frac{i{\mathfrak f}_A}
{2m}\sigma_{\mu\nu}\gamma_5\boldsymbol
\tau\partial ^\nu \boldsymbol A ^\mu (x) \} \psi (x)
\end {eqnarray}
The one meson exchange corresponds to the diagrams in
Fig. \ref{diag}.

After the application ordinary
diagram technique
\cite{lan}
we obtain the appropriate matrix element
of transition due to T-odd
$ \rho $-meson exchange:
\begin {eqnarray}
M_\rho=i\bar g_\rho \frac {g_\rho^2 \kappa} {2m} \{(\bar u_4 \gamma_\mu
\tau_j u_2) (\bar u_3 \sigma _ {\zeta\nu}
e _ {3lj} \tau_l u_1) (p_4-p_2) ^ \nu D_\rho ^ {\mu\zeta} (p_4-p_2)
\nonumber
\\ -
( \bar u_3 \gamma_\mu \tau_j u_1) (\bar u_4 \sigma _ {\zeta\nu}
e _ {3lj} \tau_l u_2) (p_4-p_2) ^ \nu D_\rho ^ {\mu\zeta} (p_4-p_2)
\nonumber
\\  -
( \bar u_4 \gamma_\mu \tau_j u_1) (\bar u_3 \sigma _ {\zeta\nu}
e _ {3lj} \tau_l u_2) (p_4-p_1) ^ \nu D_\rho ^ {\mu\zeta} (p_4-p_1)
\nonumber
\\ +
( \bar u_3 \gamma_\mu \tau_j u_2) (\bar u_4 \sigma _ {\zeta\nu}
e _ {3lj} \tau_l u_1) (p_4-p_1) ^ \nu D_\rho ^ {\mu\zeta} (p_4-p_1) \}.
\label {mtr}
\end {eqnarray}
$D ^ {\mu\nu} _ \rho (q) = \frac {q_\mu q_\nu/m^2_\rho-g _ {\mu\nu}}
 {q^2-m_\rho^2} $ represents the $\rho $-meson propagator,
$e_{ilj}$ is the antisymmetric tensor,
$\sigma_{\mu\nu}=\frac{i}{2}(\gamma_\mu\gamma_\nu-\gamma_\nu\gamma_\mu)$.
Let's substitute
$u_1\equiv u (p_1) \theta_1 $,
$u_2\equiv u (p_2) \theta_2 $ $ \ldots $
to the expression (\ref {mtr}), where
$
u (p) =
\left (
\begin {array}{c}
\sqrt {\varepsilon+m} \; \phi \\
\sqrt {\varepsilon-m} \; (\boldsymbol \sigma \boldsymbol n) \; \phi
\end {array}
\right)~,
$
$ \phi $ and $ \theta $ are spin and isospin nucleon wave functions.
Setting
$p_1=p ~, ~~~ p_2 =-p ~, ~~~ p_3=p+q ~, ~~~ p_4 =-p+q $ ~,
$$
\theta_1 =
\left (
\begin {array} {c}
1 \\
0
\end {array}
\right) ~, ~~~
\theta_2 =
\left (
\begin {array} {c}
0 \\
1
\end {array}
\right) ~, ~~~
\theta_3 =
\left (
\begin {array} {c}
1 \\
0
\end {array}
\right) ~, ~~~
\theta_4 =
\left (
\begin {array} {c}
0 \\
1
\end {array}
\right),
$$

we find the T-odd amplitude of the scattering of a proton by
a neutron due to $\rho $-meson exchange:
\begin {eqnarray}
\label{rm}
f _ {T\rho}^{pn} (\boldsymbol q) =-
\frac {M} {16\pi\varepsilon p} =
\frac {\bar g_\rho g_\rho^2\kappa (I _ {32} \boldsymbol \sigma _ {41}
\times \boldsymbol n\cdot \boldsymbol q-I _ {41} \boldsymbol
\sigma _ {32} \times \boldsymbol n \cdot \boldsymbol q)}
{ 2\pi m (m_\rho^2+4\boldsymbol p^2)} \nonumber
\\
= i\frac {\bar g_\rho g_\rho^2\kappa ((\boldsymbol n\boldsymbol
\sigma _ {31}) (\boldsymbol q\boldsymbol \sigma _ {42})- (\boldsymbol n
\boldsymbol \sigma _ {42}) (\boldsymbol q \boldsymbol \sigma _ {31}))}
{ 2\pi m (m_\rho^2+4\boldsymbol p^2)} ~,
\end {eqnarray}
where $ \boldsymbol \sigma _ {41} = \phi ^+ _4 \boldsymbol \sigma
\phi_1 $, $I _ {31} = \phi ^ + _ 3
\phi_1 $ and so on,
$\boldsymbol p$ and
$\varepsilon$
are the center of mass
nucleon momentum and the energy,
$\boldsymbol n=\frac{\boldsymbol p}
{\mid \boldsymbol p \mid} $.
The Fierz transform
$\boldsymbol \sigma_{41}I_{32}-
\boldsymbol \sigma_{32}I_{41}=i\boldsymbol \sigma_{31}\times
\boldsymbol \sigma_{42}$ was used to put the exchange amplitude to the
direct channel.
Here and everywhere further we shall use the scattering amplitude
normalized by the condition:
$ \sigma _ {tot} =4\pi \mathop{\mathrm{Im}}f (0) $.
The amplitude is derived
from ordinary amplitude (\ref{fei}) through division  by the
incident particle
wave number.
In such a normalization the amplitude
of scattering is invariant relative to the Lorenz
transform along the incident
particle momentum (small angle scattering is implied).
The
T-odd P-even $p-p$ and $n-n$ amplitudes
corresponding to
$\rho$-meson exchange is absent.
The amplitude (\ref{rm})  gives
nonrelativistic
T-odd N-N potential
\begin{equation}
V_\rho^{T}=\bar g_\rho g_\rho^2\kappa
(\tau_1\times\tau_2)_3 \boldsymbol r\times
\boldsymbol p\cdot (\boldsymbol \sigma_1-\boldsymbol \sigma_2)
\frac{m_\rho^2}{m^2}\frac{e^{-m_\rho r}}{4\pi r}
\left(\frac{1}{m_\rho r}+\frac{1}{m_\rho^2 r^2}\right).
\end{equation}
\par
The matrix element describing
$A_1 $-meson exchange is:
\begin {eqnarray}
M_A =-\frac {g_A {\mathfrak f}_A}
 {2m} \{(\bar u_4 \gamma_\mu \gamma_5 \tau_j u_2)
(\bar u_3 \sigma _ {\zeta\nu} \gamma_5
\tau_j u_1) (p_4-p_2) ^ \nu D_A ^ {\mu\zeta} (p_4-p_2) \nonumber
\\ -
( \bar u_3 \gamma_\mu \gamma_5 \tau_j u_1) (\bar u_4 \sigma _ {\zeta\nu}
\gamma_5
\tau_j u_2) (p_4-p_2) ^ \nu D_A ^ {\mu\zeta} (p_4-p_2) \nonumber
\\ -
( \bar u_4 \gamma_\mu \gamma_5\tau_j u_1) (\bar u_3 \sigma _ {\zeta\nu}
\gamma_5
\tau_j u_2) (p_4-p_1) ^ \nu D_A ^ {\mu\zeta} (p_4-p_1) \nonumber
\\ +
( \bar u_3 \gamma_\mu \gamma_5 \tau_j u_2) (\bar u_4 \sigma _ {\zeta\nu}
\gamma_5
\tau_j u_1) (p_4-p_1) ^ \nu D_A ^ {\mu\zeta} (p_4-p_1) \}.
\label {ma}
\end {eqnarray}
After similar calculations we obtain
\begin {eqnarray}
\label{app}
f _ {TA}^{pp} (q) =
\frac {i g_A {\mathfrak f}_A} {\pi m} \frac {{\boldsymbol p} ^2} {m_A^2 (m_A^2+4
{\boldsymbol p} ^2)} ((\boldsymbol \sigma _ {42} \boldsymbol p)
(\boldsymbol \sigma _ {31} \boldsymbol q)
+ (\boldsymbol \sigma _ {42} \boldsymbol q) (\boldsymbol \sigma _ {31}
\boldsymbol p))
\end {eqnarray}
for the $p-p $ scattering and
\begin {eqnarray}
\label{apn}
f _ {TA}^{pn} (\boldsymbol q) =
\frac {-i g_A {\mathfrak f}_A} {4\pi m}
\frac {3m_A^2+4 {\boldsymbol p} ^2} {m_A^2 (m_A^2+4 {\boldsymbol p} ^2)}
((\boldsymbol \sigma _ {42} \boldsymbol p) (\boldsymbol \sigma _ {31}
\boldsymbol q)
+ (\boldsymbol \sigma _ {42} \boldsymbol q) (\boldsymbol \sigma _ {31}
\boldsymbol p))
\end {eqnarray}
for the $p-n $ scattering.
The corresponding potential is
\begin{eqnarray}
V_A^T=g_A{\mathfrak f}_A\frac{m_A^2}{m^2}
((\boldsymbol\sigma_1 \boldsymbol p)(\boldsymbol \sigma_2
\boldsymbol r)+(\boldsymbol \sigma_2
\boldsymbol p)(\boldsymbol \sigma_1 \boldsymbol r)-
(\boldsymbol \sigma_1\boldsymbol \sigma_2)
(\boldsymbol p\boldsymbol r))
\nonumber \\
\times(\boldsymbol \tau_1 \boldsymbol \tau_2)
\frac{e^{-m_A r}}{8\pi r}
\left(\frac{1}{m_A r}+\frac{1}{m_A^2 r^2}\right)
+h.c.~~~.
\end{eqnarray}
The T-even constants are approximately equal to
$g_\rho=2.79 ~ $, $g_A =\frac {5} {3\sqrt {2}} g_\rho $,
$\kappa=3.7$ \cite{sim}.
For T-odd constants the following restrictions were derived:
$ | \bar g_\rho | < 6.7\times 10 ^ {-3} ~ $,
$ |{\mathfrak f}_A | < 3\times 10 ^ {-5} $ \cite{sim}.

\section{T-odd P-even forward $p-d$ elastic scattering
amplitude}

For the nucleon-deuteron scattering we have from (\ref{fei}):
\begin {eqnarray}
\label {f20}
F (0) = \frac {i} {2\pi} \int Sp _ {\sigma_1 \sigma_2}
\biggl \{
\bigl(
\Gamma_1 (\boldsymbol b-\boldsymbol b_1) +
\Gamma_2 (\boldsymbol b-\boldsymbol b_2)
~~~~~~~~~~~~~~~~~~~~~~~~~~
\nonumber \\
-
\Gamma_1 (\boldsymbol b-\boldsymbol b_1) \Gamma_2
(\boldsymbol b-\boldsymbol b_2)
\theta (z_1-z_2)-
\Gamma_2 (\boldsymbol b-\boldsymbol b_2)
\Gamma_1 (\boldsymbol b-\boldsymbol b_1) \theta (z_2-z_1)
\bigr)
\nonumber \\
\times\delta (\boldsymbol r_1-\boldsymbol r_2) \rho (\boldsymbol r_1)
\biggr \}
d^3
\boldsymbol r_1d^3\boldsymbol r_2
d^2
\boldsymbol b~,
\end {eqnarray}
where $\theta(z)$ is the step-function and
$\boldsymbol r_\alpha\equiv \{\boldsymbol b_\alpha,z_\alpha \}$.
We again use the amplitudes
normalized by the condition
$\sigma_{tot}=4\pi \mathop{\mathrm{Im}}F(0)$.
The
double scattering terms of the
expression (\ref{f20}) imply that
the incident particle spin wave function
is acted by
$\Gamma_1$, and then by  $\Gamma_2$
if the particle firstly
strikes
with the nucleon 1. If the
first collision occurs with
nucleon 2 the $\Gamma_2$ acts firstly.

The
profile-function $\Gamma
(\boldsymbol b)$ is connected with the
$N-N$ nucleon scattering amplitude by
the
relation:
$f(\boldsymbol q)=\frac{i}{2\pi} \int \Gamma(\boldsymbol b)
e^{-i\boldsymbol q\boldsymbol b} d^2\boldsymbol b $.
If we rewrite the equation
(\ref{f20})
in terms of amplitudes and form factors
$G^{(+)}(\boldsymbol q)=\int
\left\{ \int_0 ^
{+\infty} \rho (\boldsymbol r) dz
\right\}e^{i{\bf qb}}d^2\boldsymbol b~$,
$G^{(-)}(\boldsymbol q)=\int
\left\{  \int _ {-\infty} ^ {0}
\rho (\boldsymbol r) dz
\right\}e^{i{\bf qb}}d^2\boldsymbol b~$,
$\boldsymbol r\equiv \{\boldsymbol b,z\}$
we can see that in the expression derived
\begin {eqnarray}
F (0) =Sp_{\sigma_1 \sigma_2} \biggl \{(f_1 (0) +f_2 (0)) G (0)
\biggr \}~~~~~~~~~~~~~~~~~~~~~~~~~~~~~
\nonumber \\ +
\frac {i} {2\pi
} Sp _ {\sigma_1 \sigma_2} \biggl \{\int (f_1 (-\boldsymbol q)
f_2 (\boldsymbol q) G ^ {(+)} (2\boldsymbol q) +
f_2 (\boldsymbol q) f_1 (-\boldsymbol q) G ^ {(-)}
( 2\boldsymbol q)) d^2\boldsymbol q \biggl \}
\nonumber
\end {eqnarray}
the area of integration
over the momentum transferred $ \boldsymbol q $ is restricted by the
deuteron
form factor, which dependence on the
momentum transferred is sharper than that for $N-N$ scattering amplitude.
It allows one to take out the $N-N$ amplitude from the
integral or, that is equivalent, to use
the formal
profile-function giving the correct expression for the $N-N$ amplitude
in a
vicinity of small angles:
\begin {eqnarray}
\Gamma_\alpha (\boldsymbol b) = \frac {2\pi} {i} \biggl \{a_\alpha +
v_\alpha (\boldsymbol \sigma \boldsymbol \sigma_\alpha) +
e_\alpha (\boldsymbol \sigma_\alpha \boldsymbol n) (\boldsymbol \sigma
\boldsymbol n)
-\frac {c_\alpha} {m} (\boldsymbol \sigma + \boldsymbol \sigma_\alpha)
\times \boldsymbol n \cdot\frac {
\partial} {\partial \boldsymbol b} \nonumber \\-
\frac {d_\alpha} {m^2} (\boldsymbol \sigma_\alpha
\frac {\partial} {\partial \boldsymbol b}) (\boldsymbol \sigma
\frac {\partial
} {\partial \boldsymbol b})
-\frac {ih_\alpha} {m} (\boldsymbol \sigma_\alpha \boldsymbol n)
(\boldsymbol \sigma
\frac {\partial} {\partial \boldsymbol b}
) - \frac {ih ^\prime_\alpha} {m} (\boldsymbol \sigma \boldsymbol n)
(\boldsymbol \sigma_\alpha
\frac {\partial} {\partial \boldsymbol b})
\biggr \}\delta^{(2)} (\boldsymbol b).
\label{expr}
\end {eqnarray}
Here $ \delta^{(2)} (\boldsymbol b) $ is two demensional
delta-function, $m $ is the nucleon mass.
Although
the approximation is very rough, especially for the spin-dependent
form factors $G$ \cite{cher},
 we use
it to simplify calculations.
The expression (\ref{expr})
corresponds to the scattering amplitude of the incident
proton by the $\alpha$  deuteron nucleon
($\alpha=1$ denotes the p-p scattering
and $\alpha=2$ denotes the $p-n$ one)
with constant
$a_\alpha, v_\alpha, e_\alpha, c_\alpha, d_\alpha, h_\alpha,
h^\prime_\alpha$:
\begin {eqnarray}
f_\alpha (\boldsymbol q) =
a_\alpha + v (\boldsymbol \sigma_\alpha \boldsymbol \sigma) + e
(\boldsymbol \sigma_\alpha \boldsymbol n) (\boldsymbol \sigma
\boldsymbol n) + \frac {ic_\alpha} {m} ( \boldsymbol \sigma_\alpha
+\boldsymbol \sigma)\cdot \boldsymbol q \times \boldsymbol n
\nonumber \\
+ \frac {d_\alpha}
{m^2} (\boldsymbol \sigma_\alpha \boldsymbol q) (\boldsymbol \sigma
\boldsymbol q) + \frac {ih_\alpha} {m} ( \boldsymbol
\sigma_\alpha \boldsymbol n) (\boldsymbol \sigma \boldsymbol q) +
\frac {ih^\prime_\alpha} {m} (\boldsymbol \sigma \boldsymbol n)
(\boldsymbol \sigma_\alpha \boldsymbol q)~.
\nonumber
\end {eqnarray}
It follows from (\ref{rm}), (\ref{apn}), (\ref{app})
that $\rho$-meson
exchange gives
\begin{equation} h_1^\rho=h_1^{\rho\prime}=0,~~~~~~
h_2^\rho=-h_2^{\rho\prime}=-\frac{i\bar g_\rho g_\rho^2
\kappa}{2\pi(m_\rho^2+4p^2)}~,
\end{equation}
and $A_1$-meson exchange gives
\begin{equation}
h_1^A=h_1^{A\prime}=\frac{ig_A{\mathfrak f}_Ap^2}{\pi m_A^2(m_A^2+4p^2)}~,~~
h_2^A=h_2^{A\prime}=-\frac{ig_A{\mathfrak f}_A(3m_A^2+4p^2)}{4\pi m_A^2(m_A^2+4p^2)}.
\end{equation}

Let us represent the nucleon density at deuteron $\rho
(\boldsymbol r)$ as:
\begin{eqnarray}
\rho (\boldsymbol r)=\frac 14 \bigl\{
A_0+A_1\boldsymbol S\left( \boldsymbol \sigma _1+
\boldsymbol \sigma _2\right)
+A_2(\boldsymbol S\boldsymbol r)(\boldsymbol \sigma _1+
\boldsymbol \sigma _2)\cdot \boldsymbol r+
A_3(\boldsymbol S\boldsymbol r)^2
\nonumber
\\
+A_4(\boldsymbol \sigma_1
\boldsymbol \sigma _2)
+A_5(\boldsymbol \sigma_1\boldsymbol r)(\boldsymbol \sigma_2
\boldsymbol r)+
A_6((\boldsymbol \sigma _
1\boldsymbol S)(
\boldsymbol \sigma_2\boldsymbol S)+(
\boldsymbol \sigma
_2\boldsymbol S)(\boldsymbol \sigma _1
\boldsymbol S))
\nonumber
\\
+A_7(\boldsymbol \sigma _1
\boldsymbol r)(\boldsymbol \sigma _2
\boldsymbol r)(\boldsymbol S\boldsymbol r)^2+
A_8(\boldsymbol \sigma _1\boldsymbol \sigma _2)
(\boldsymbol S\boldsymbol r)^2+
A_9((\boldsymbol \sigma _1\times \boldsymbol S
\cdot \boldsymbol r)
\nonumber
\\
*(\boldsymbol \sigma
_2\times \boldsymbol S\cdot \boldsymbol r)
+(\boldsymbol \sigma _2\times \boldsymbol S\cdot \boldsymbol r)
(\boldsymbol \sigma _1\times \boldsymbol S\cdot \boldsymbol r))+
A_{10}((\boldsymbol \sigma _1\boldsymbol r)
(\boldsymbol \sigma _2\boldsymbol S)
(\boldsymbol S\boldsymbol r)
\nonumber
\\
+(\boldsymbol \sigma _1
\boldsymbol S)(\boldsymbol \sigma _2\boldsymbol r)
(\boldsymbol S\boldsymbol r)
+
(\boldsymbol \sigma
_1\boldsymbol r)(\boldsymbol S
\boldsymbol r)(\boldsymbol \sigma _2\boldsymbol S)+
(\boldsymbol S\boldsymbol r)(\boldsymbol \sigma _2
\boldsymbol r)(\boldsymbol \sigma _1\boldsymbol S))
\nonumber
\\
\label{rho}+
T_0(\boldsymbol \sigma_1 \times\boldsymbol \sigma_2\cdot\boldsymbol S)+
T_1((\boldsymbol \sigma _1\boldsymbol r)
(\boldsymbol \sigma _2\times \boldsymbol S\cdot \boldsymbol r)
+ (\boldsymbol \sigma _2\boldsymbol r) (\boldsymbol
 \sigma _1\times \boldsymbol S\cdot \boldsymbol r))
\\+
T_2 ((\boldsymbol S\boldsymbol r) ((\boldsymbol
\sigma _1 + \boldsymbol \sigma _2) \times \boldsymbol S
\cdot \boldsymbol r) + \nonumber ((\boldsymbol
\sigma _1 +\boldsymbol \sigma _2) \times \boldsymbol S
\cdot \boldsymbol r) (\boldsymbol S\boldsymbol r))\bigr\}.
\nonumber
\end {eqnarray}
The density is simultaneously the spin
density matrix of the both deuteron nucleons. The operator of the deuteron
spin $\boldsymbol S$ is a parameter describing deuteron orientation.
To find density for the concrete deuteron orientation we must
take matrix element
of $\rho(\boldsymbol r)$
over deuteron spin state.
The terms $T_0,T_1$ and $T_2$ are the T-odd terms.
The real functions $A_0 ... A_{10} $ can be found from the
deuteron wave function.
\[
\phi _{m}=\frac{1}{\sqrt{4\pi }}\left( \frac{U(r)}{r}+\frac{1}{\sqrt{8}}
\frac{W(r)}{r}S_{12}\right) \chi _{m},
\]
where $U(r)$ is the radial deuteron S-wave function and $W(r)$ is the radial
D-function, $\chi _{m}$ is the spin wave function of two nucleons
with the spin $m$,
$S_{12}=3(\boldsymbol \sigma _{1}\boldsymbol n)
(\boldsymbol \sigma _{2}\boldsymbol n)-(\boldsymbol \sigma _{1}
\boldsymbol \sigma _{2})$. Taking
into account that $\rho (r)=8\phi (2r)\phi ^{+}(2r)$
we can
find the nucleon density matrix
for the deuteron being at the state with the spin projection 1:
\[
\langle 1 \mid \rho (r))\mid 1\rangle = \frac{1}{\pi }\left(
\frac{U(2r)}{r}
+\frac{W(2r)}{\sqrt{8}r}S_{12}\right) \chi _{1}\chi _{1}^{+}
\left(
\frac{U(2r)}{r}+\frac{W(2r)}{\sqrt{8}r}S_{12}\right)
\]
From the other side,
$\langle 1\mid \rho \mid 1\rangle $
can be obtained by taking the matrix
element
from expression (\ref{rho}) over the state with the spin projection 1.
By comparing these expressions
we find:

$A_0(r)=u^2-8uw+16w^2,\;\;\;A_1=u^2-2uw-8w^2,\;\;\;r^2A_2=6uw+12w^2,$
\medskip
 $r^2A_3=12uw-12w^2,\;\;\;\;\;\;\;A_4=8w^2+8uw-u^2,\,\,\,\,\,\,\,\,\,\,r^2A_5 =-24w^2, $
\medskip
 $A_6=u^2-2uw+4w^2,\;\;\;\;\;\;\;r^4A_7=72w^2,\;\;\;\;\;\;\;\;\;\;\;\;r^2A_8 =-12w^2, $
\medskip
 $r^2A_9=-6uw,\;\;\;\;\;\;\;\;\;\;\;\;\;\;\;\;\;\;\;\;r^2A_{10}=-12w^2.$

The functions $u$ and $w$ are connected with $
U $ and $W$ by: $W(2r)=\sqrt{8\pi }rw(r)$, $U(2r)=\sqrt{\pi }ru(r)$.
Substituting the expressions for $\rho(\boldsymbol r)$ and
for the profile-function to the equation (\ref
{f20}) it is possible to calculate the T-odd P-even $p-d$ forward scattering
amplitude.
\begin {eqnarray}
\label {form}
 F_T(0)= \pi i ((\boldsymbol \sigma \cdot \boldsymbol S\times
\boldsymbol n)
 (\boldsymbol S\boldsymbol n) + (\boldsymbol S\boldsymbol n)
(\boldsymbol \sigma \cdot \boldsymbol S \times \boldsymbol n))~~~~~~~~~~~~
\nonumber
\\ \times
\int\biggl\{
\{(a_1v_2+v_1a_2) z^2T_2-\frac 1 {4m^2} (a_1d_2+d_1a_2) (zT_2 ^{\prime}
\nonumber \\  +T_2) -\frac {c_1c_2} {2m^2} (zT_2 ^ {\prime} -3T_2)
+ \frac i {2k} (c_1v_2+c_2v_1+c_1e_2+c_2e_1) zT_2
\nonumber \\
-\frac 1{2k}(h_1v_2+h_2v_1)(z^3A_7+zA_8+2zA_{10})+\frac 1 {2k}
(h_1 ^ {\prime} v_2+h_2 ^ {\prime} v_1) (zA_8
\nonumber \\
+zA_9+3zA _
{10}) -\frac i {4m^2} (c_1h_2+c_2h_1) (\frac {A_6 ^ {\prime}}
z-z^2A_7-3A_9+zA _ {10} ^ {\prime}
\nonumber \\  - A _ {10})
+ \frac 1{2m}(h_1e_2+h_2e_1)(z^3A_7+zA_8+2zA_{10})\}\biggr\}dz~.
\end {eqnarray}
Everywhere in (\ref {form}) $A_n, T_n $ are the functions of $z $,
the prime means the derivative with respect to $z$.
The
two different contributions to the T-odd scattering amplitude
exist. The first one is due to T-odd N-N scattering
taking part in
the double scattering of the incident proton by the
deuteron nucleons.
For spherical deuteron $A_2,A_3,A_5,A_7,A_8,A_9,A_{10}$ are equal to
zero and the only term proportional to the
$(c_1h_2+c_2h_1) \frac {A_6 ^ {\prime}} z$ remains.
In any case the spin-dependent N-N
scattering amplitude
is needed to generate
T-odd effect in the p-d amplitude.
The spin-dependent N-N amplitude decreases with the energy, so the
T-odd p-d amplitude decreases too. We don't speak here about possible
short range T-odd N-N forces which, in principle, can grow with the
energy and compensate decreasing of the spin-dependent N-N amplitude.

\begin{figure}[h]
\hspace{0.0cm}
\vspace{0.0cm}
\epsfxsize =9.3 cm
\epsfbox[0 90 400 400]{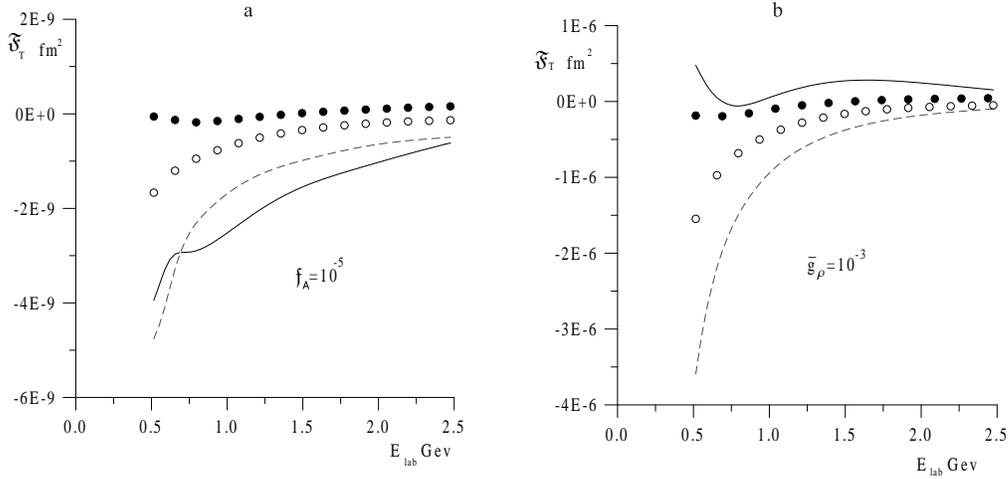}
\vspace{0.0cm}
\caption{T-odd P-even p-d forward elastic scattering
amplitude calculated with T-odd P-even N-N amplitude
due to $A_1$-meson exchange (a) and due to $\rho$-meson
exchange (b). Solid curve is real part and dashed curve is
imaginary part. Black and empty circles correspond  to the real and
imaginary parts respectively calculated for "spherical" deuteron.}
\label{mes9}
\end{figure}
\begin{figure}[h]
\hspace{2.0cm}
\vspace{0.0cm}
\epsfxsize =5. cm
\epsfbox[0 0 400 700]{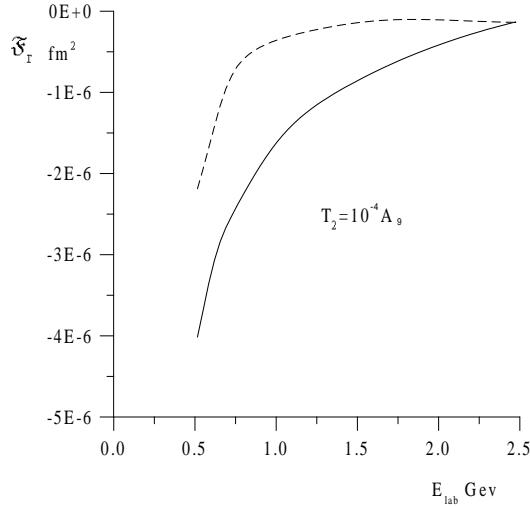}
\vspace{-2.0cm}
\caption{T-odd P-even p-d forward elastic scattering
amplitude calculated with T-odd P-even
impurity in the nucleon density at the deuteron.
Solid curve is real part and dashed curve is
imaginary part.}
\label{den}
\end{figure}

The second contribution is due to T-odd impurity in the
nucleon density at deuteron.
We can see from the expression (\ref{form}) that  only the $T_2$ term
gives rise to the p-d amplitude.  $T_1$ and $T_0$ terms do not
contribute to the p-d amplitude because they contain less than $2$
deuteron spin operators.
 The T-even  spin dependent p-d scattering amplitude was considered in
\cite{cher}.

\section{Results}

The T-even amplitudes $a, b, c... $ are taken from
the SAID phase shift analysis \cite{said} and the
deuteron S-,D- radial functions are taken from \cite{deu}.
Setting T-odd constants equal
$\bar g_\rho=10 ^ {-3} $ and ${\mathfrak f}_A=10 ^ {-5} $ we obtain
results shown in Fig. \ref{mes9}a
The restriction on ${\mathfrak f}_A$ is more rigorous due to strict contribution
of the $A_1$-exchange
mechanism
to the neutron dielectric moment and the
accuracy $10^{-9}-10^{-10}~fm^2$ for amplitude measurements or
$10^{-7}-10^{-8}~mb$ for
T-odd cross section measurements is needed
to find ${\mathfrak f}_A= 10^{-5}$.
The restriction for $\bar g_\rho$ is more gentle and requires
the accuracy $10^{-6}-10^{-7}~fm^2$  for the amplitude (Fig. \ref{mes9}b) and
$10^{-4}-10^{-5}~mb$
for the cross section measurements.
We see that deuteron non sphericity makes results more
optimistic.

For $T_2$ impurity to the density we take
$T_2=10^{-4}A_9$ because the T-even $A_9$ term has similar
structure as $T_2$.
To test the impurity at this level we
need the accuracy $10^{-6}-10^{-7}~fm^2$
and $10^{-4}-10^{-5}~mb$
 for the amplitude (Fig. \ref{den}) and cross section
measurements
respectively.

So, the accuracy of TRV collaboration ($10^{-6}~ mb$) will allow to
obtain new constraints for ${\mathfrak f}_A$ and $\bar g_\rho$. Note,
that the real part of the T-odd P-even forward amplitude can be measured
in the spin rotation \cite{barr} experiment.

The author is grateful to the Prof. V. G. Baryshevsky for
the remark concerning the
importance of a deuteron non sphericity and
to the Dr. K. G. Batrakov and Dr.  I. Ya. Dubovskaya for discussions and remarks.

\end {document}